\documentstyle[11pt,newpasp,epsf]{article}

\begin{document}

\title{Energy transport, overshoot, and mixing in the atmospheres of
very cool stars}

\author{Hans-G\"unter Ludwig}
\affil{Lund Observatory, Box 43, 22100 Lund, Sweden; hgl@astro.lu.se}

\begin{abstract}
We constructed hydrodynamical model atmospheres for mid M-type main-,
as well as pre-main-sequence objects. Despite the complex chemistry
encountered in such cool atmospheres a reasonably accurate
representation of the radiative transfer is possible. The detailed
treatment of the interplay between radiation and convection in the
hydrodynamical models allows to study processes usually not accessible
within the framework conventional model atmospheres. In particular, we
determined the efficiency of the convective energy transport, and the
efficiency of mixing by convective overshoot. The convective transport
efficiency expressed in terms of an equivalent mixing-length parameter
amounts to values around $\approx 2$ in the optically thick, and
$\approx 2.8$ in the optically thin regime. The thermal structure of
the formally convectively stable layers is little affected by
convective overshoot and wave heating, i.e. stays close to radiative
equilibrium. Mixing by convective overshoot shows an exponential
decline with geometrical distance from the Schwarzschild stability
boundary. The scale height of the decline varies with gravitational
acceleration roughly as $g^{-\frac{1}{2}}$, with 0.5 pressure scale
heights at \mbox{$\log \mathrm{g}$}=5.0.
\end{abstract}


\section{Introduction}

The increasing number of late M-type stars, brown dwarfs, and
extrasolar planets discovered by infrared surveys and radial velocity
searches has spawned a great deal of interest in the atmospheric
physics of these objects. Their atmospheres are substantially cooler
than e.g. the solar atmosphere, allowing the formation of molecules,
or even liquid and solid condensates. Convection is a ubiquitous
process in these atmospheres shaping the thermal structure and the
distribution of material.  Hydrodynamical simulations of solar and
stellar granulation including a realistic description of radiative
transfer have become an increasingly powerful and handy instrument for
studying the influence of convective flows on the the structure of
late-type stellar atmospheres as well as on the formation of their
spectra. Here we report on efforts to construct hydrodynamical models
for mid M-type atmospheres. We consider this as an intermediate step
on the way to model brown dwarf and planetary atmospheres. The
motivation was twofold: Firstly, pre-main-sequence evolutionary models
of M-dwarfs and brown dwarfs based on mixing-length theory (MLT)
depend sensitively on the mixing-length parameter~\mbox{$\alpha_{\mathrm{MLT}}$} (Baraffe et
al. 2002). Secondly, the distribution of dust clouds in brown dwarfs
depends on the efficiency of mixing by convective
overshoot. Conventionally, convection is described within the
simplistic picture of MLT dependent on free parameters. Our
hydrodynamical models provide a description essentially from first
principles free of the uncertainties of MLT, putting stellar models on
a firmer footing.

\section{Model overview}

\begin{figure}
\vspace{-1em}
\begin{center}
\epsfysize=0.33\textheight
\epsfbox{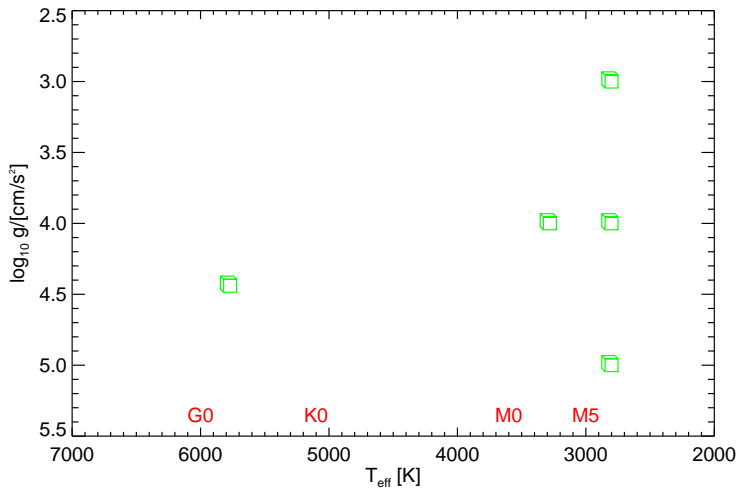}
\end{center}
\vspace{-2em}
\caption{The radiation-hydrodynamics models in the effective
temperature-gravity plane (cubes). The approximate spectral class is
indicated at the temperature axis.}
\label{f:modover}
\end{figure}

Figure~\ref{f:modover} illustrates in the location of the
hydrodynamical model atmospheres considered in this investigation in
the \mbox{$\mathrm{T}_{\mathrm{eff}}$}-\mbox{$\log \mathrm{g}$}-plane: three M-type models are located at
\mbox{$\mathrm{T}_{\mathrm{eff}}$}=2800\,K with \mbox{$\log \mathrm{g}$}=3.0, 4.0, and 5.0 which form a
\mbox{$\log \mathrm{g}$}-sequence.  To assess the temperature effects a somewhat hotter
M-type model is located at \mbox{$\mathrm{T}_{\mathrm{eff}}$}=3280\,K and \mbox{$\log \mathrm{g}$}=4.0. For comparison
we also considered a solar model at \mbox{$\mathrm{T}_{\mathrm{eff}}$}=5777\,K, \mbox{$\log \mathrm{g}$}=4.44. All
models possess solar chemical composition.

The RHD simulations were performed with a convection code developed by
{\AA}.~ Nordlund and. R.F.~Stein (see Stein \&\ Nordlund 1998, and
references therein).  The code implements a consistent treatment of
compressible gas flows together with non-local radiative energy
exchange in three spatial dimensions. The radiative transfer is
treated in LTE approximation, the wavelength dependence of the
radiation field is represented by a small number of wavelength
bins. Open lower and upper boundaries, as well as periodic lateral
boundaries are assumed. The effective temperature of a model (i.e. the
average emergent radiative flux) is controlled indirectly by
prescribing the entropy of inflowing material at the lower
boundary. Magnetic fields are neglected.  Opacities and
equation-of-state have been adapted to the conditions encountered in
the M-type atmospheres. In particular, the equation-of-state accounts
for $\mathrm{H}_2$ molecule formation, the opacities include
contributions of molecular lines. The opacities were extracted from
the opacity data base of the PHOENIX model atmosphere code (for a
description of the code see Hauschildt \&\ Baron 1999). The
application of the Nordlund \&\ Stein convection code to M-type
objects in described in more detail in Ludwig, Allard, \&\ Hauschildt
(2002, hereafter LAH).

\begin{figure}[!tb]
\vspace{-1em}
\begin{center}
\epsfysize=0.4\textheight
\epsfbox{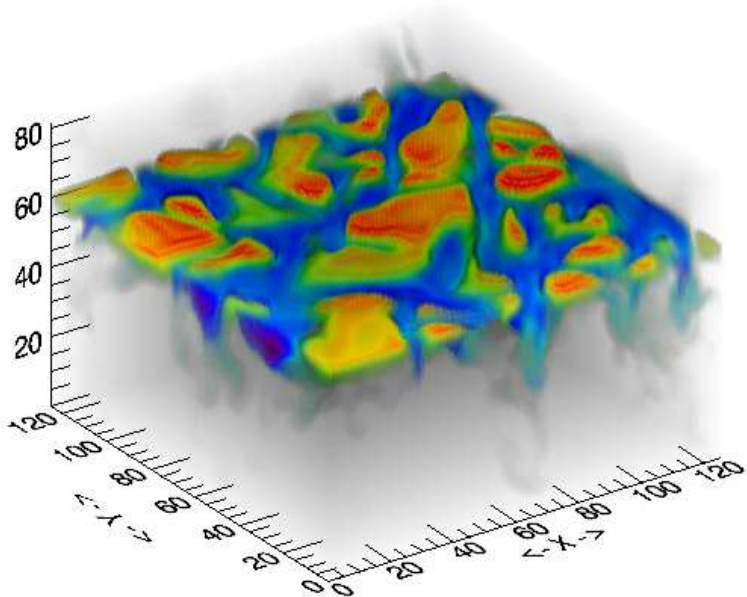}
\end{center}
\vspace{-2em}
\caption{Typical temperature field during the evolution of a 3D
hydrodynamical model of a late-type atmosphere. Lighter shades of gray
correspond to higher temperatures.}
\label{f:approach}
\end{figure}

Figure~\ref{f:approach} visualises the flow structure (here via the
temperature field) generically encountered in our hydrodynamical
models for late-type stars. The computational domain is a Cartesian
box representing a small volume at the stellar surface.  It comprises
the optically thin photospheric layers as well the uppermost layers of
the convection zone which extends deeper into the star. A granulation
pattern is visible in layers around optical depth
unity. Concentrated, plume-like downdrafts dominate the
sub-photospheric flow, the optically thin layers are harbouring a
mixture of overshooting and wave motions with associated temperature
fluctuations.

\begin{figure}[!tb]
\vspace{-1em}
\begin{center}
\epsfysize=0.4\textheight
\epsfbox{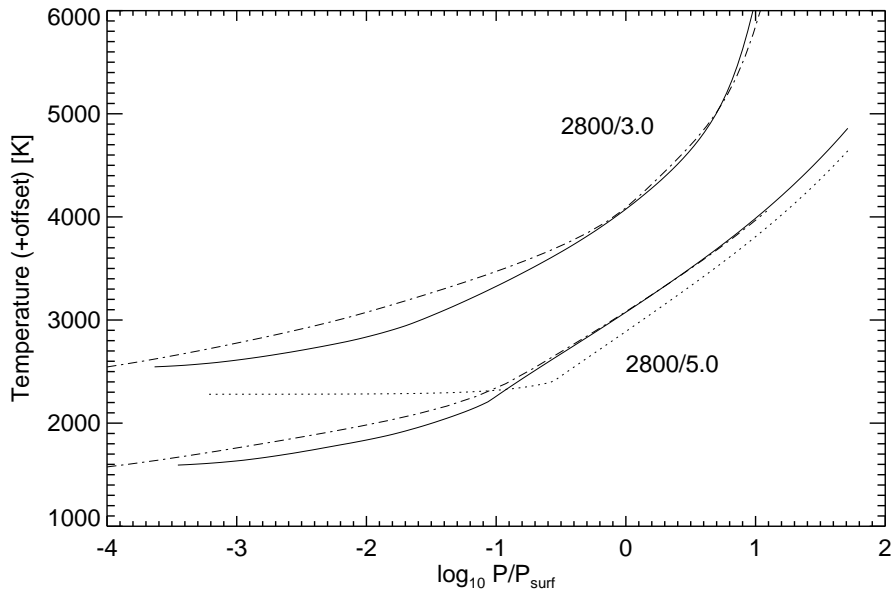}
\end{center}
\vspace{-2em}
\caption{Comparison of the pressure-temperature structure between 1D
hydrostatic model atmospheres in radiative-convective equilibrium
based on the OBM (solid lines) with 4 wavelength groups, and PHOENIX
models based on direct opacity sampling employing several 10\,000
wavelength points (dash-dotted). Shown are two examples with
\mbox{$\mathrm{T}_{\mathrm{eff}}$}=2800\,K and \mbox{$\log \mathrm{g}$}=3.0 and 5.0, respectively. For clarity, the
\mbox{$\log \mathrm{g}$}=3.0 models were shifted by +1000\,K. Also shown is a model
employing grey radiative transfer (dotted line). The pressure is given
in units of the pressure at Rosseland optical depth
unity~$P_{\mathrm{surf}}$.}
\label{f:obmmdwarfs}
\end{figure}

We want to derive quantitative estimates of the mixing by convective
overshoot, as well as obtain a measure of the efficiency of the
convective energy transport. For addressing these issues the RHD
models have to give a reasonably accurate representation of the actual
atmospheric conditions. Here we are particularly concerned about the
radiative energy transport, which is complicated by the huge number of
molecular absorption lines. In our hydrodynamical models we use a
multigroup technique (dubbed {\it Opacity Binning Method\/}, hereafter
OBM) for modelling the radiative energy exchange which employs four
groups for representing the wavelength dependence of the radiation
field (see LAH for details). The wavelength groups have been optimised
for the \mbox{$\mathrm{T}_{\mathrm{eff}}$}=2800\,K and \mbox{$\log \mathrm{g}$}=5.0 model. Figure~\ref{f:obmmdwarfs}
illustrates the accuracy which is achieved with the OBM. While there
are differences between the atmospheric structure based on the OBM and
PHOENIX models, which are based on direct opacity sampling, the OBM
provides a significant improvement with respect to the simple grey
approximation. Differences increase as one chooses atmospheric
parameters away from the ones for which the OBM was optimised. We did
not generate specifically optimised groupings for all atmospheres
under consideration, since we were interested to study the systematic
change of model properties with effective temperature and
gravitational acceleration. I.e it appeared advantageous to leave the
input physics the same in all models, only optimised for one
temperature and gravity.

%\begin{figure}
%\plotone{ludwighgf8.eps}

To mitigate the effects of the still present shortcomings in the
radiative transfer we took a {\em differential approach\/} when
measuring model properties: whenever possible we compared
hydrodynamical and hydrostatic model atmospheres based on the same
radiative transfer scheme, i.e. here the OBM.

\section{Granulation in M-type objects}

\begin{figure}[!tb]
\plotone{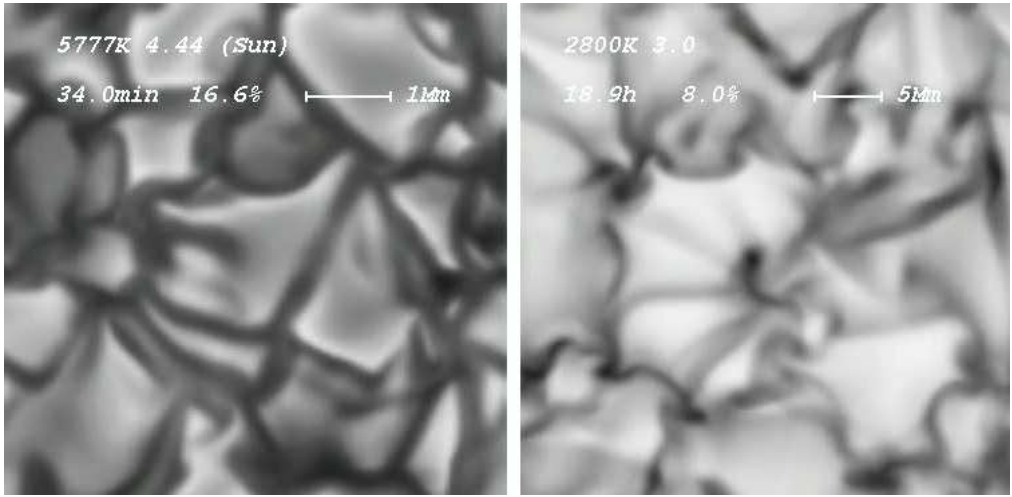}
\caption{Granulation pattern in the solar model (left panel) and the
\mbox{$\mathrm{T}_{\mathrm{eff}}$}=2800\,K, \mbox{$\log \mathrm{g}$}=3.0 M-type model (right panel). Shown is the
emergent intensity in the continuum. At the particular instants in
time the relative intensity contrasts amount to 16.6\,\% in the solar
case, and 8.0\,\% in the M-type case. Note, the different spatial
scales.}
\label{f:grancomp}
\end{figure}

Our small sample of hydrodynamical M-type atmospheres gives some
insight into the properties of stellar granulation at cooler
temperatures. Figure~\ref{f:grancomp} compares the granulation pattern
of the \mbox{$\mathrm{T}_{\mathrm{eff}}$}=2800\,K, \mbox{$\log \mathrm{g}$}=3.0 model with the one of a solar model.
The first thing to note is that surface convection in M-type objects
produces a granular pattern qualitatively resembling solar
granulation: bright extended regions of upwelling material which are
surrounded by dark concentrated lanes of downflowing material. The
dark lanes form an interconnected network.  Looking more closely,
granules are less regularly delineated in M-type objects, the
inter-granular lanes show a higher degree of variability in terms of
their strength. A feature which is uncommon in the solar granulation
pattern are the dark ``knots'' found in or attached to the
inter-granular lanes. The knots are associated with strong downdrafts
which carry a significant vertical component of angular momentum. The
width of the inter-granular lanes to the typical granular size is
smaller in our M-type objects. Inspecting the velocity field (not
shown) in vicinity of the continuum forming layers shows less
pronounced size differences. This indicates that the relatively
broader lanes in the solar case are the result of a stronger smoothing
of the temperature field due to a more intense radiative energy
exchange, i.e. the effective Pecl{\'e}t number of the flow is larger
around optical depth unity in the cooler objects.

\section{Mixing by atmospheric overshoot}

Stellar atmospheres around $\mbox{$\mathrm{T}_{\mathrm{eff}}$}\approx 2800\,\mathrm{K}$ are too hot
to allow for a significant formation of dust grains. However, already
at slightly cooler effective temperatures, grain formation sets in and
dust grains become major opacity contributors, i.e. an important
factor in determining the thermal structure of the atmosphere. In
fact, the spectral energy distribution in the range of effective
temperatures $2500\,\mathrm{K} \leq T \leq 1500\,\mathrm{K}$ is crucially linked
to the distribution of dust grains in the atmosphere (e.g. Allard,
Hauschildt, \&\ Schwenke 2000). The amount of dust which is present is
determined by chemical condensation and evaporation processes as
sources and sinks, as well as macroscopic transport processes which
carry dust grains away from their sites of formation. In M-type and
cooler atmospheres the transport is dominated by two opposing
processes: gravitational settling of dust grains and their mixing due
to the presence of velocity fields, either related to convection or
global circulations induced by rotation. In our models, no dust
formation takes place but we nevertheless find it worthwhile to give a
characterisation of the atmospheric mixing due to convection and
convective overshoot. This can give at least a first order
approximation of how convective mixing might operate when dust is
actually present.

Formally, we are interested in a statistical representation of the
mean transport properties of the convective velocity field in vertical
direction. The horizontal advection of dust grains by the convective
velocity field probably produces horizontal inhomogeneities in the
dust distribution. We neglect these here since i) we are targeting at
the application of our results in standard 1D stellar atmosphere
models, ii) the horizontal inhomogeneities are small scale (the size
of a convective cell), i.e. hardly observable and iii) the
uncertainties in our understanding of the dust formation process
itself perhaps limits the achievable accuracy anyway. In view of the
last point we present a proxy of the convective mixing only, without
trying to derive a detailed statistical description of the transport
properties of the convective velocity field, i.e.  extracting its
effective transport coefficients (c.f. Miesch, Brandenburg, \&\
Zweibel 2000, for an example of a more stringent treatment).

As proxy measuring the mixing properties of the velocity field we
introduce the mixing velocity $v_{\mathrm{mix}}$
\begin{equation}
v_{\mathrm{mix}} \equiv \frac{F}{\langle\rho\rangle}
\end{equation}
($\rho$ denotes the mass density, $\langle.\rangle$ temporal and
horizontal averaging) where $F$ is the temporally and horizontally
averaged vertical gross mass flux
\begin{equation}
F \equiv \langle\rho u_{\mathrm{z}}\rangle_{\mathrm{t,up}}
\end{equation}
($u_{\mathrm{z}}$ denotes the vertical velocity). For $F$ the average
is taken over surface areas with upwards directed flow only. Note,
that the average total mass flux vanishes. $F$ measures the amount of
mass flowing in and out a certain layer. If the in- and outflow
happens sufficiently disordered, this mass exchange goes hand in hand
with mixing of material.

As we shall see below $v_{\mathrm{mix}}$ exhibits an exponential
height dependence. With such an dependence a characteristic mixing
frequency~$f_{\mathrm{mix}}$ can be written as
\begin{equation}
f_{\mathrm{mix}} \equiv -\frac{1}{\langle\rho\rangle}\frac{dF}{dz}
 = v_{\mathrm{mix}} \left( \frac{1}{H_{\mathrm{v}}}+\frac{1}{H_\rho}\right),
\end{equation}
where $H_{\mathrm{v}}$ denotes the scale height of $v_{\mathrm{mix}}$, and
$H_\rho$ the density scale height.

\subsection{Subsonic filtering}

As mentioned previously, the atmospheric velocity field is a
superposition of advective, overshooting motions and acoustic waves
generated by convection in deeper layers (see also Ludwig \&\
Nordlund 2000). The wave motions contribute little if at all to the
mixing due to their spatially coherent, oscillatory character. The
overshooting, convective motions tend to decay with distance from the
Schwarzschild stability boundary, while the wave motions tend to
increase in amplitude due to the exponential decrease of the density
with increasing height. This leads to the situation that beyond a
certain height the atmospheric velocity field is even dominated by
wave motions. In order to get a reliable estimate of the mixing it is
therefore necessary to remove the wave contributions to the velocity
field before evaluating the gross mass flux~$F$.

\begin{figure}[!tb]
\vspace{-1em}
\begin{center}
\epsfysize=0.4\textheight
\epsfbox{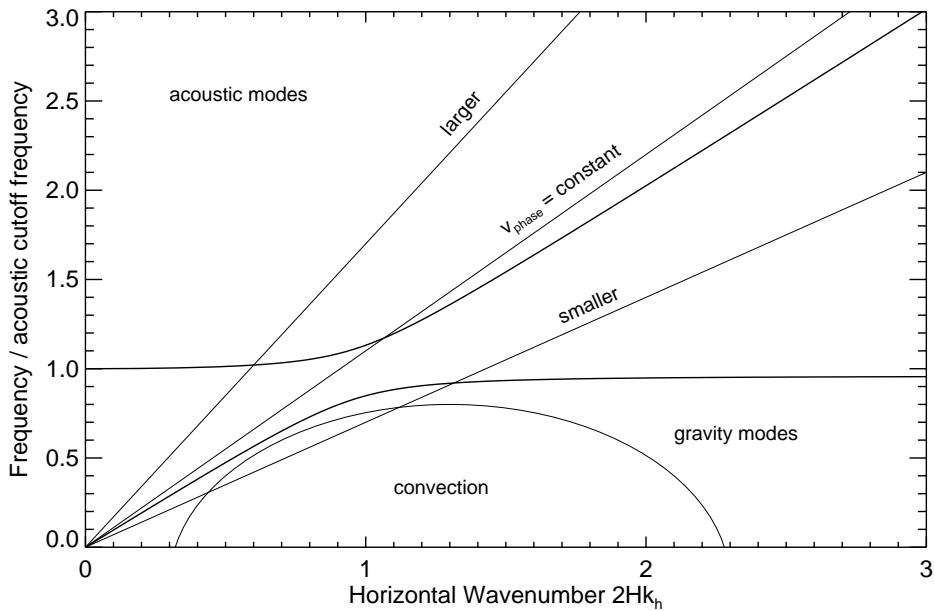}
\end{center}
\vspace{-2em}
\caption{Schematic illustration of subsonic filtering: Only components
in wavenumber-frequency domain below a prescribed phase
speed~$v_{\mathrm{phase}}$ are retained. They preferentially belong to
convective motions.}
\label{f:subsonic}
\end{figure}

We removed the wave contributions by subsonic filtering --- a
technique invented in the context of solar observations for cleaning
images from ``noise'' stemming from the solar 5\,minute oscillations
(Title et al. 1989). Figure~\ref{f:subsonic} schematically illustrates
this filtering technique.  In short, one considers a time sequence of
images and removes features with horizontal phase
speeds~$v_{\mathrm{phase}}$ greater than a prescribed threshold. This
is achieved by Fourier filtering of spatial-temporal data in the
$k$-$\omega$ domain.  For every depth layer in our data cubes we
performed a 3D Fourier analysis (one temporal, two spatial dimensions)
of the vertical mass flux retaining only contributions below a preset
velocity threshold. In practice, acoustic and convective contributions
are not as cleanly separated as shown in the Fig.~\ref{f:subsonic},
and one must find the right balance between removing as much acoustic
components as possible while retaining as much as possible convective
contributions. We always studied a sequence of phase speed thresholds
in order to judge the success of the procedure.

%\begin{figure}
%\plotone{ludwighgf4.eps}

\subsection{Mixing in the solar atmosphere}

\begin{figure}[!tb]
\vspace{-1em}
\begin{center}
\epsfysize=0.4\textheight
\epsfbox{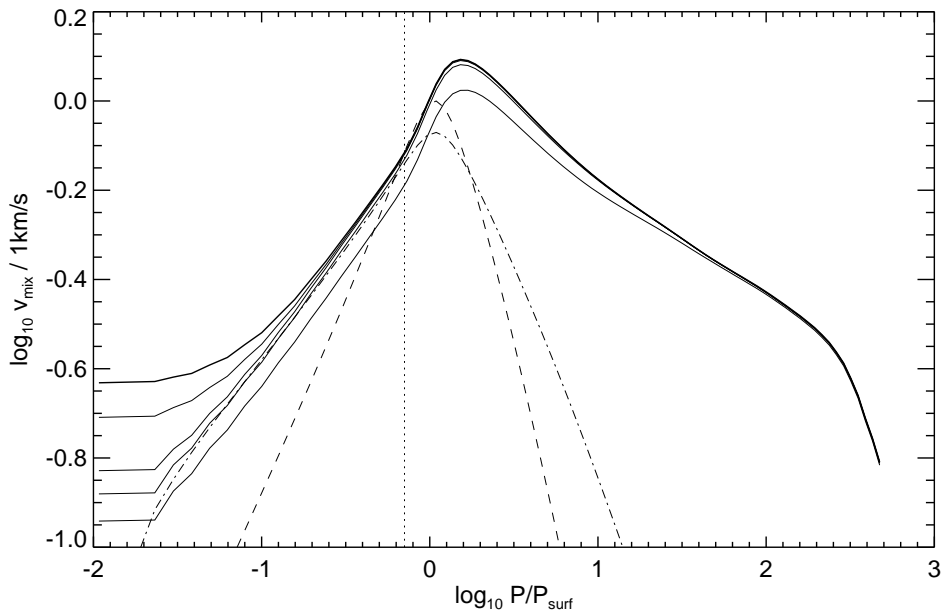}
\end{center}
\vspace{-2em}
\caption{Mixing velocity in a solar model as a function of pressure:
the unfiltered data (thick solid line) were subsonically filtered with
$v_{\mathrm{phase}}$ = 12, 6, 3, and 1.5\,km/s (thin solid lines from
top to bottom). The approximate location of the Schwarzschild boundary
of convective stability is indicated by the dotted line. The velocity
profiles of convective eigenmodes with a horizontal wavelength of
1.5\,Mm (dashed line) and 5.0\,Mm (dash-dotted line) are also shown. The
pressure is given in units of the pressure at Rosseland optical depth
unity~$P_{\mathrm{surf}}$. The velocity plateau at the upper boundary
is a numerical artifact of the upper boundary condition of the
hydrodynamical model.}
\label{f:filters}
\end{figure}

Despite we are not primarily interested in the Sun here, it is
worthwhile to look at the Sun as a reference object to compare with.
Figure~\ref{f:filters} shows $v_{\mathrm{mix}}$ in our solar model for
various degrees of subsonic filtering. It is clearly visible that the
subsonic filtering has the strongest impact on $v_{\mathrm{mix}}$ in
the uppermost atmospheric layers. As advertised before {\em
$v_{\mathrm{mix}}$ exhibits an exponential decline with height ($\log
P \propto z$) after appropriate subsonic filtering\/}. This feature
was put forward by Freytag, Ludwig, \&\ Steffen (1996) as generic
feature of convective overshoot. Moreover, we see that too low a
velocity threshold removes also convective features, as visible by the
reduction of the velocity in the deeper, convection dominated
layers. Quantitatively, for the Sun we find that
$H_{\mathrm{v}}=(1.9\pm 0.2) H_{\mathrm{p}}$, where $H_{\mathrm{p}}$
denotes the local pressure scale height.

The exponential decline of $v_{\mathrm{mix}}$ is also motivated from
studying linear convective modes. In Fig.~\ref{f:filters} we plotted
the velocity profiles of to linear convective eigenmodes with
horizontal wavelength of 1.5 and 5.0\,Mm, respectively. We used the
temporally and horizontally averaged hydrodynamical model as
background mode. The absolute velocity amplitude of the modes has been
scaled to match $v_{\mathrm{mix}}$ from the hydrodynamical model. We
see that the velocity field of the modes exhibit an exponential
``leaking'' into the formally convective stable layers, and the
decline is more rapid for the mode with shorter wavelength. We
further see that the mode with 5\,Mm wavelength shows a decline
similar to the decline of $v_{\mathrm{mix}}$. Interestingly, the
wavelength of the fitting mode is significantly larger than the
horizontal scale of the dominant convective structures on the Sun ---
the granules with typical sizes of around 1.5\,Mm.

%\begin{figure}
%\plotone{ludwighgf5.eps}

\subsection{Mixing in M-type atmospheres}

\begin{figure}[!tb]
\vspace{-1em}
\begin{center}
\epsfysize=0.4\textheight
\epsfbox{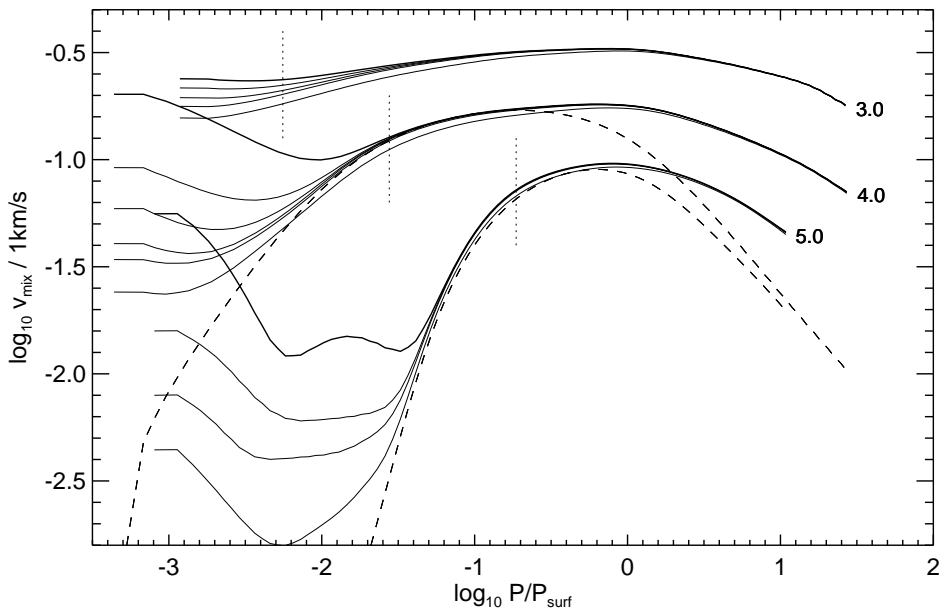}
\end{center}
\vspace{-2em}
\caption{Mixing velocities in the M-type models with \mbox{$\mathrm{T}_{\mathrm{eff}}$}=2800\,K and
\mbox{$\log \mathrm{g}$}=3.0, 4.0, and 5.0 (groups of solid lines, top to bottom) as a
function of pressure: shown are the unfiltered data (thick solid
lines) and subsonically filtered data (thin solid lines, top to
bottom) with $v_{\mathrm{phase}}$= 8.0, 4.0, 2.0, 1.0\,km/s
(\mbox{$\log \mathrm{g}$}=3.0 and \mbox{$\log \mathrm{g}$}=4.0); 1.0, 0.5, 0.25\,km/s (\mbox{$\log \mathrm{g}$}=5.0). The
approximate locations of the Schwarzschild boundaries are indicated by
the dotted lines. The dashed lines depict velocity profiles of
convective modes with horizontal wavelength of 3.0\,Mm (\mbox{$\log \mathrm{g}$}=4.0) and
250\,km (\mbox{$\log \mathrm{g}$}=5.0).}
\label{f:filterm}
\end{figure}

Figure~\ref{f:filterm} shows $v_{\mathrm{mix}}$ for the M-type models
at \mbox{$\mathrm{T}_{\mathrm{eff}}$}=2800\,K. With decreasing gravity the zone of convective
instability extends further and further into the optically thin
atmosphere, leaving little room for overshoot in the lowest gravity
model. Reading off an exponential decline rate is very uncertain here.
However, we find a slow decline with $H_{\mathrm{v}}\approx 6
H_{\mathrm{p}}$. For the \mbox{$\log \mathrm{g}$}=4.0 model we find a value close to
solar of $1.7\,H_{\mathrm{p}}$, for the \mbox{$\log \mathrm{g}$}=5.0 model
$0.5\,H_{\mathrm{p}}$. Qualitatively, overshooting is less pronounced
in models of higher gravity. This is of course related to the fact
that the buoyancy forces scale proportional to gravity, making
buoyancy more effective in confining the convective motions to the
formally unstable regions. Empirically we find that the scale height
of $v_{\mathrm{mix}}$ roughly scales as $g^{-\frac{1}{2}}$.

Trying to match the velocity profile in the overshooting regions worked
only partially. For the lower gravity models the fit is not very good
which might be related the situation that convection reaches very
heigh atmospheric layers. We oriented the horizontal wavelength of the
linear modes at the largest sizes of structures the computational box
could accommodate in the respective models.

%\begin{figure}
%\plotone{ludwighgf6.eps}

\section{The efficiency of the convective energy transport}

\begin{figure}[!tb]
\vspace{-1em}
\begin{center}
\epsfysize=0.4\textheight
\epsfbox{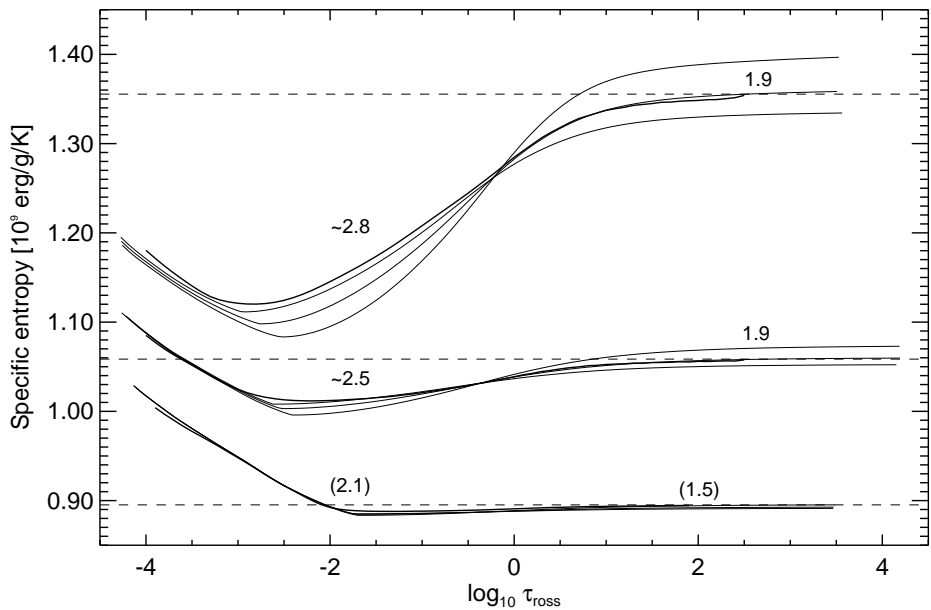}
\end{center}
\vspace{-2em}
\caption{Entropy structure of the hydrodynamical models (thick solid
lines; \mbox{$\log \mathrm{g}$}=3.0, 4.0, 5.0, top to bottom) in comparison to standard
mixing-length models (thin solid lines). For each hydrodynamical model
three MLT models are plotted with \mbox{$\alpha_{\mathrm{MLT}}$}=1.5, 2.0, and 2.5. The dashed
lines depict an extrapolation towards the asymptotic entropy
encountered deep in the convective zone. Numbers indicate approximate
mixing-length parameters necessary to match the hydrodynamical
structure with standard models.}
\label{f:entropy}
\end{figure}

Convection is an important energy transport mechanism in M-type
stars. In standard model atmospheres it is treated in the framework of
MLT. The question is whether the simplistic MLT is actually capable to
provide a reasonable description of the convective energy transport
under conditions encountered in the M-type atmospheres.
Figure~\ref{f:entropy} shows a comparison of the thermal structure of
the hydrodynamical model atmospheres and standard 1D hydrostatic model
atmospheres in radiative-convective equilibrium assuming
different mixing-length parameters. The hydrodynamical models have
been averaged temporally and horizontally on surfaces of constant
optical depth. This procedure ensures a particularly good preservation
of the energy transport properties of the hydrodynamical models
(Steffen, Ludwig, \&\ Freytag 1995). Besides the mixing-length
parameter MLT contains a number of further ``hidden'' parameters
intrinsic to the specific formulation of MLT which was chosen. We
emphasise that a well-defined calibration of the mixing-length
parameter must be always given with reference to the specific
formulation. Here, we are using the formulation as given by Mihalas
(1978).

The sensitivity of the structure of the standard models to the
mixing-length parameter increases with decreasing gravity. This is in
line with the earlier statement that stellar structure models depend
sensitively on the mixing-length parameter at lower gravities. Towards
the main sequence the sensitivity is largely reduced making a precise
calibration of \mbox{$\alpha_{\mathrm{MLT}}$}\ difficult --- but of course also less
important. The M-type atmospheres offer the remarkable opportunity to
study convection also under optically thin conditions. We roughly
separate the models into a optically thin and thick part and derive
the corresponding mixing-length parameters suitable to fit the
hydrodynamical structures in the different regions. The mixing-length
parameter of the optically thick region is most relevant for global
stellar structure models, while the one of the optically thin part is
most relevant for stellar atmospheres.

%\begin{figure}
%\plotone{ludwighgf7.eps}

The entropy gradients of hydrodynamical and hydrostatic models
correspond closely in the highest, convectively stable layers (with
$\frac{ds}{d\tau}<0$). By construction the standard models are in
radiative equilibrium in these layers. This indicates that in the
hydrodynamical models overshooting and wave heating are not
particularly efficient, i.e. in the hydrodynamical models these layers
are also close to radiative equilibrium. In the convectively unstable
layers (with $\frac{ds}{d\tau}>0$) the situation is markedly
different. The stratifications (except for the \mbox{$\log \mathrm{g}$}=5.0 model which
is close to convective equilibrium, i.e. almost adiabatically
stratified) are neither close to radiative nor convective
equilibrium. The detailed balance between radiative and convective
energy transport determines the thermal structure. This makes the
stratification sensitive to the values of the mixing-length parameter
which provides a measure of the efficiency of the convective energy
transport. As indicated in Fig.~\ref{f:entropy} we find a
mixing-length parameter between 2.1 and 2.8 in the optically thin
regions, and between 1.5 and 1.9 in the optically thick region. The
extreme values of 1.5 and 2.1 are found in the \mbox{$\log \mathrm{g}$}=5.0 model and are
somewhat uncertain due to low sensitivity to changes of the
mixing-length parameter. The \mbox{$\mathrm{T}_{\mathrm{eff}}$}=3280\,K and \mbox{$\log \mathrm{g}$}=4.0 model (not
plotted) shows values of 2.8 (optically thin part) and 2.2 (optically
thick part).

Besides the detailed quantitative results two aspects are apparent:
Firstly, it is not possible the specify a unique mixing-length
parameter fitting the hydrodynamical stratification over the whole
depth range. This illustrates principal shortcomings in the
MLT. Secondly, the derived mixing-length parameters are falling in the
range commonly considered. If one does not have high demands on the
accuracy MLT provides a reasonable scaling of the energy transport
efficiency with a fixed --- perhaps solar calibrated --- mixing-length
parameter in the regime of M-type atmospheres.

\section{Final remarks}

While our results do not apply to the brown dwarf regime directly, it
is tempting to extrapolate the mixing properties we found in our mid
M-type objects. Our coolest and highest gravity model points towards a
modest extra mixing by convective overshoot in brown dwarf
atmospheres. The mixing would stay confined to layers in vicinity of
the Schwarzschild stability boundary, and would not lead to a complete
distribution of dust in the atmosphere (see LAH for more details).

Mixing-length theory does perform reasonably well in M-type
atmospheres. To get a better quantitative description one might try to
calibrate beside \mbox{$\alpha_{\mathrm{MLT}}$}\ the ``internal'' parameters of MLT. M-type
atmospheres appear particularly well suited for this undertaking since
convection takes place under optically thick and thin conditions.

We reiterate that our results apply to atmospheres of solar
metallicity. We expect a markedly different outcome for metal poor
atmospheres.

\acknowledgments
 
This project benefitted from financial support of the Walter
Gyllenberg Foundation, and the Swedish Vetenskapsr{\aa}det.

\end{document}